\documentstyle{article}

\title{Wetting on Lines and Lattices of Cylinders}
\author{W. R. Osborn and J. M. Yeomans \\
	Theoretical Physics, 1 Keble Road, Oxford OX1 3NP}

\begin{document}
\maketitle
\abstract{This paper discusses wetting and capillary condensation
transitions on a line and a rectangular array of cylinders using an
interface potential formalism. For a line of cylinders, there is a
capillary condensation transition followed by complete wetting if the
cylinders are sufficiently close together. Both transitions disappear
as the cylinder separation is increased. The dependence of the wetting
phase diagram of a rectangular array of cylinders is discussed as a
function of the chemical potential, substrate--fluid interaction
strength and surface tension.

}

\pagebreak
\section{Introduction}
The wetting of planar surfaces is now well understood
\cite{Schick,SullivanTelodeGamma,Dietrich}. However, in
many realistic situations, substrates are far from planar. A
particularly important example is provided by porous
media, whose wetting properties have implications for fluid flow
\cite{Wilkinson}, oil recovery \cite{Hinch} and also the probing of the
fractal geometry of surfaces \cite{PfeiferCole}. Our aim in
this paper is to describe wetting and capillary
condensation on lines and arrays of cylinders as a step towards
understanding the properties of binary fluids in complicated geometries.

For such complicated geometries rigorous theoretical methods quickly
become intractable and approximations must be made. A profitable
approach has been to use an interface
potential which replaces the density profile at a fluid--fluid
interface by a sharp kink and uses a local surface tension
\cite{Schick,Dietrich}. This
approach is valid far from the bulk critical point and for wetting
layers thicker than a few intermolecular spacings. It has the
advantage that it is easily implemented, yet provides qualitatively
correct phase diagrams.

Cheng and Cole \cite{ChengCole} and Napi\'{o}rkowski et al.\
\cite{Napiorkowski} applied the interface
potential approach to wetting in a corner, Darbellay and
Yeomans \cite{Darbellay} to wetting in a slit, and Robbins et al.\
\cite{Robbins} to wetting on a line of slits. Dobbs et al.\
\cite{Harvey1} extended the approach to treat two spheres, and
subsequently a square array of cylinders \cite{Harvey3} in both the
grand canonical and the more physically realistic canonical ensembles.
In each case, sensible qualitative results were obtained for the phase
behaviour although, as one might expect, subtle details of the
interface position are not given correctly \cite{Dietrich}.

In Section 2 we use such an interface potential approach to study
wetting on a line of cylinders. We consider the case of Van der Waals
interparticle interactions. If the cylinders are
close enough together for a capillary condensation transition to
occur, the system undergoes complete wetting as $\widetilde{\Delta\mu}
\rightarrow 0$. Otherwise, the cylinders behave individually rather
than collectively and the wetting transition is suppressed.

In Section 3, a rectangular array of cylinders is considered. The
phase diagram is determined as a function of the aspect ratio of the
array, the chemical potential, the strength of the Van der Waals
interactions, and the surface tension. Limiting cases in which this
substrate reverts to the line of cylinders described in Section 2 and
the square array considered by Dobbs and Yeomans \cite{Harvey3} are
discussed.

Although using the interface potential approach significantly
simplifies the problem, it is still not easy to study the more
complicated geometries which will model a porous medium more closely.
For example, even for an array of spherical substrates, the non-linear
differential equations that must be solved become two- rather than
one-dimensional. Therefore, in Section 4, we describe a simpler way of
modelling the different phases, which is amenable to extension to
more complicated regular and random geometries. A comparison to the
solutions of the interface potential approach allows us to assess its
usefulness.

\nopagebreak
Our results are summarised in Section 5.

\section{An Infinite Line of Cylinders}
Firstly, we consider wetting on an infinite line of identical
cylinders lying along the $x$-axis with their axes parallel to the
$z$-axis, as shown in Figure~ 1. The cylinders have radius $r_{0}$ and
their separation is $L'$. The relevant part of the grand potential per
cylinder, per unit length in the $z$-direction, is taken to be

\begin{equation}
\Phi = 4 \left( \int_{0}^{+L'/2} \left\{\sigma\sqrt{1 + l_{x}^{2}(x)} +
\widetilde{\Delta\mu} \left( l(x) - \frac{\pi r_{0}^{2}}{2 L'} \right)
\right\} dx + W[l(x)] \right)
\label{granpot}
\end{equation}
where $l(x)$ is the interface position and the subscript $x$ denotes
differentiation with respect to $x$.

The first term in equation (\ref{granpot}) is the free energy of the
liquid--gas interface, the surface tension $\sigma$ multiplied by the surface
area. The second is a bulk term due to the excess cost of the adsorbed,
unfavourable liquid. If $\rho_{l}$ and $\rho_{g}$ are the liquid and
gas number densities, respectively, then the free energy per unit volume of
the liquid phase over and above that of the gas phase is

\begin{equation}
\widetilde{\Delta\mu} = (\mu_{c} - \mu^{*})(\rho_l-\rho_g)
\end{equation}
with $\mu^{*}$ the chemical potential of the fluid in the system and
$\mu_{c}$ the chemical potential at bulk liquid--gas coexistence.

The final term models the interparticle interactions which, for
non-retarded Van-der-Waals forces, can be written

\begin{equation}
W[l(x)] = \int_{0}^{+L'/2}\int_{l(x)}^{\infty} \Pi[x',y(x'),r_{0},L']\,dy\,dx'
\label{W[l]}
\end{equation}
with a disjoining pressure

\begin{equation}
\Pi (x,y,r_{0},L') = \sum_{i}
\int_{\mbox{\scriptsize cylinder $i$}} \frac{W_{0}'}{\left|\underline{r} -
\underline{r}^{\prime}\right|^{6}} \, d\underline{r}^{\prime}
\label{Pi}
\end{equation}
where the summation is taken over all cylinders $i$.
Note that the integrals in equations (\ref{W[l]}) and (\ref{Pi}) are
over the gas and substrate; all other interactions are either
independent of $l(x)$ or can be reformulated as integrals over these
regions. The strength of the interactions is $W'_{0} = A /\pi^{2}$
where $A$ is the conventionally defined Hamaker constant.

The integral in equation (\ref{Pi}) cannot be performed analytically, in
contrast to the cases of a spherical substrate \cite{GelfandLipowsky} and a
cylindrical pore \cite{Cole1}, but the numerical
result is well fit by a function

\begin{eqnarray}
\Pi[x,y,r_{0},L'] & \approx & W_{0}' \sum_{i=-\infty}^{\infty}  \left(
\frac{\pi e^{-(l_{i}/r_{0}-1)}} {6(l_{i}-r_{0})^{3}} + \frac{3 \pi^{2}
r_{0}^{2}}{8 l_{i}^{5}} \right) \label{D.P.}
\end{eqnarray}
where $l_{i}$ is the distance from the centre of cylinder $i$ to the
point $\underline{r}$. The first term in the
expression (\ref{D.P.}) is accurate at small distances from a
cylinder, when the substrate acts like a flat plane. The second gives
the correct behaviour in the long distance limit. The fit
(\ref{D.P.}) agrees with numerical integration of the
disjoining pressure to within 10\%, with the largest discrepancy
occuring in the crossover region, at a distance of about $r_{0}$ from
the surface of a cylinder. This discrepancy is unimportant because the
contribution of the interactions to the total free energy is small
compared to that of the surface energy at this distance.

The grand free energy, $\Phi$, can be
minimised with respect to $l(x)$ using the Euler-Lagrange formula,
yielding a non-linear differential equation

\begin{equation}
\frac{d}{dx}\left( \frac{\sigma l_{x}}{\sqrt{(1+l_{x}^{2})}}\right) -
\widetilde{\Delta\mu} + \Pi \left[ x,l(x),r_{0},L' \right] = 0
\label{ELmin}
\end{equation}
A solution where the liquid forms bridges between the cylinders (see
Figure~ 1) may be found if this is solved with boundary conditions
$l_{x} = 0$ at $x = 0$ and $x = L'/2$.

To find the unbridged solution, where the interface wraps around each
individual cylinder, it is necessary to formulate the
problem in polar coordinates with the origin at the centre of a
cylinder. The Euler-Lagrange minimisation then gives

\begin{equation}
\frac{d}{d\theta}\left(
\frac{\sigma l_{\theta}}{\sqrt{(l^{2}+l_{\theta}^{2})}} \right) -
\widetilde{\Delta\mu} \; l(\theta) + \Pi \left[
\theta,l(\theta),r_{0},L' \right] \, l(\theta) -
\frac{\sigma l(\theta)}{\sqrt{(l^{2}+l_{\theta}^{2}})} = 0
\label{ELmin2}
\end{equation}
The boundary conditions are $l_{\theta} = 0$ at $\theta = 0$ and
$\theta = \pi/2$, where $l_{\theta} \equiv dl(\theta)/d\theta$.


The solutions to the differential equations (\ref{ELmin}) and
(\ref{ELmin2}) can be found numerically using a relaxation method for
different values of the parameters
$\mu = \widetilde{\Delta\mu} \, r_{0} / \sigma$, $L = L'/r_{0}$ and
$W_{0} = W_{0}' / (\sigma r_{0}^{2} )$. Once the interface
profiles are known, the grand free energy of each phase can be
calculated from equation (\ref{granpot}) using numerical integration
techniques, allowing comparison of the free energies and determination
of the stable configuration.

The resulting phase diagram is shown in Figure 2.
For  $L < 2$ the cylinders are
overlapping and the problem is not defined. The unbridged phase
is stable at large $\mu$, as expected. For $2<L<L_{c}(W_{0})$, as
$\mu$ is decreased there is a first-order phase
transition to the bridged phase and then, as $\mu
\rightarrow 0$ the interface unbinds to infinity, an example of a
complete wetting transition. As the surface unbinds to large
distances, it becomes flat and it follows from
(\ref{D.P.}) and (\ref{ELmin}) that

\begin{equation}
\widetilde{\Delta\mu} \sim \frac{\pi^{2} r_{0}^{2} \, W_{0}'}{2 \, L'
\, l^{4}} \mbox{ as } l \rightarrow \infty
\end{equation}
This shows that the line of cylinders is behaving, as expected, like a
plate of effective thickness $\pi r_{0}^{2} / L'$.

For $L > L_{c}$, capillary condensation does not occur as
$\mu$ decreases and the complete wetting transition
is suppressed by the substrate geometry. For vanishingly small
$W_{0}$ at $\mu = 0$, the surface area is the only
relevant quantity. The liquid--gas surface area per unit length for a
single cylinder in the unbridged phase is $2 \pi r_{0}$ while that of
a bridged film is $2 L_{c} r_{0}$, giving $L_{c}(0) = \pi$ in
agreement with the numerical solution.

As $W_{0}$ is increased at fixed $\mu$, the interfaces lie further
from the substrates. For an approximately flat, bridged interface
the surface energy is virtually unchanged by this.
However, for the unbridged solution the surface area
and hence the surface energy must increase as the interface moves.
Thus the unbridged solution becomes less
favourable for a given $\mu$, as seen in Figure~ 2.

\section{An Infinite Rectangular Array of Cylinders}
An infinite number of lines of cylinders can be brought together to produce a
rectangular array with inter-line distance $D'$, as shown in Figure
3. As $D = D'/r_{0}$ is
reduced, this system shows a crossover from the behaviour of a line of
cylinders to that reported in Dobbs and Yeomans \cite{Harvey3} for a
square array.

Two new phases might be expected to exist. The first of these consists
of bridging between lines of cylinders as well as between the cylinders
in one line. To find such a doubly-bridged solution it is necessary to
use a polar coordinate system centred on an interstitial site such as
point $A$ in Figure~ 3. The Euler-Lagrange minimisation then gives

\begin{equation}
\frac{d}{d\theta}\left(
\frac{\sigma l_{\theta}}{\sqrt{(l^{2}+l_{\theta}^{2})}} \right) +
\widetilde{\Delta\mu} \; l(\theta) - \Pi \left[
\theta',l(\theta'),r_{0},L' \right] \, l(\theta) -
\frac{\sigma l(\theta)}{\sqrt{l^{2}+l_{\theta}^{2}}} = 0
\label{ELmin3}
\end{equation}
where $l'$ and $\theta'$ are the distance and angle to point
$(l,\theta)$ from the centre of one of the four nearest cylinders. The
boundary conditions are that $l_{\theta} = 0$ at $\theta = 0$ and
$\theta = \pi/2$. Equation (\ref{ELmin3}) is solved numerically, as before.

There is also a phase where the space between the cylinders is
completely filled with liquid. The free energy of this per cylinder,
per unit length in the z-direction is
\begin{equation}
\Phi_{\mbox{\scriptsize full}} = \widetilde{\Delta\mu}( L' D' - \pi
r_{0}^{2})
\end{equation}

{}From comparisons of the free energies, the stable
configuration can be found for given values of the four parameters
$L, D, \mu = \widetilde{\Delta\mu} r_{0}/\sigma$ and $W_{0} = W_{0}'/\sigma
r_{0}^{2}$. Typical values of the Hamaker constant for a condensed
phase are of order $(0.4 - 4)\times 10^{-19}$J, while typical
interfacial energies lie in the range $(2 - 8)\times 10^{-2}\mbox{J
m}^{-2}$ \cite{Israelachvili}. If we suppose that the cylinders are of
the same size as the grains in a porous rock, then
$r_{0} \sim (1-100)\times 10^{-6}$m, giving $W_{0} \sim 10^{-6}$.

Phase diagrams are plotted in Figures 4 and 5 and are discussed below.

\begin{itemize}
\item Small L

Figure 4 shows a cross-section through the phase
diagram for $L = 2.2$ and $W_{0} = 2.5
\times 10^{-6}$. For any finite $D$, the behaviour is no longer that of
many separate horizontal lines of cylinders. Complete wetting at
$\mu = 0$ is replaced by a transition to a full phase at $\mu > 0$. For
large $D$, this transition lies along the line $D=2/\mu$. As $D$
decreases, the phase which is singly-bridged along the
$\underline{\hat{x}}$-direction becomes unstable, as expected, and the
doubly-bridged solution is favoured.
For $D=L=2.2$, symmetry demands that singly-bridged solutions cannot be
stable and we recover, as expected, a phase diagram topologically
similar to that of Dobbs and Yeomans for a square array \cite{Harvey3}.

For $D<L$, as $\mu$ is decreased the system jumps from being unbridged
to singly-bridged along $\underline{\hat{y}}$ to doubly-bridged to
full, the order of increasing liquid volume.

\item Increasing L

For $L = 2.6$, $W_{0} = 2.5 \times 10^{-6}$, the phase diagram is that
shown in Figure 5. As $L$ is increased, the quadruple point {\em B} in Figure
4 moves to lower values of $\mu$ until it coincides with point {\em
A}, when the doubly-bridged phase disappears from the phase diagram.
Moreover, the phase which is singly-bridged along
$\underline{\hat{x}}$ is stable only at increasingly high values of
$D$ as L is increased further until $L \equiv L^{*}_{c}$, when it
becomes thermodynamically unstable for all $D$. $L^{*}_{c}$ is, as
expected, approximately equal to $L_{c}$, the critical value above
which a line of cylinders does not undergo complete wetting, with
small corrections due to the influence of the other cylinders in the array.

\item Increasing $W_{0}$

When $W_{0}$ is increased, at a given value of $\mu$ the free energy
of phases with more interface close to a
surface will increase relatively more than those with less. Hence the
transition lines move to greater values of $\mu$. For $W_{0} = 2.5
\times 10^{-5}$ and $L=2.2$, the phase diagram is like that in Figure
4 but point $B$ moves to $\mu \approx 6.67$ while point
$A$ moves to values of $\mu$ and $D$ greater by $\sim 3\%$.

\end{itemize}

\section{Simple Model}
The interface potential approach already involves considerable
approximation. However, it still relies on the high symmetry of the
system considered to produce tractable, one-dimensional, non-linear
differential equations. To better model porous media it would
be desirable to be able to treat more complicated substrate geometries.
Thus, we now consider a much simpler way of modelling the phases in
the cylindrical array and compare the resulting phase diagrams with
those obtained from the interface potential approach. We find only small
discrepancies for physically realistic $W_{0}$, giving confidence that
the simpler model will give qualitatively correct phase diagrams for
more realistic models of porous media.

The approach is to approximate the interface shapes in the different
thermodynamically stable phases by simple curves that can be handled
analytically. The effect of the Van der Waals interaction is
incorporated by assuming that where an interface wraps around a
cylinder it lies a distance $r$ from the centre of the cylinder, where
$r$ follows from the flat plane result
\begin{equation}
r-r_{0} = \sqrt[3]{\frac{W_{0}' \pi}{6 \widetilde{\Delta \mu}}}
\end{equation}
Where the liquid forms bridges between cylinders, the bridges are
taken to have a radius of curvature $a = \sigma/\widetilde{\Delta \mu}
\equiv r_{0} / \mu$, which follows from minimising the free energy of a
bridge with respect to $a$. By assuming that the arcs of radius $r$
and $a$ meet tangentially, the interface shape is completely defined.

The different phases can be modelled as shown in Figure 6.

\begin{itemize}
\item Unbridged phase

A circle of radius $r$, centred on a cylinder.

\item Singly-bridged phase

An arc of radius $r$, centred on a cylinder, for $\theta > \theta_{0}$, with
$\theta$ measured from the direction of bridging. An arc of radius
$a$ tangential to this at $\theta = \theta_{0}$ , with $\frac{\partial
l}{\partial x} = 0$ at $x = L'/2$, where $\cos \theta_{0} =
\frac{L'}{2 \left(r + a\right)}$.

\item Doubly-bridged phase

An arc of radius $r$, centred on a cylinder, for $\theta_{0} < \theta
< \theta_{1}$. Two arcs of radius $a$ tangential
to this, the first at $\theta = \theta_{0}$ with $\frac{\partial
l}{\partial x} = 0$ at $x = L'/2$, where $\cos \theta_{0} =
\frac{L'}{2 \left(r + a\right)}$ and the second at
$\theta = \theta_{1}$ with $\frac{\partial l}{\partial y} = 0$ at
$y = D'/2$, where $\cos \theta_{1} = \frac{D'}{2 \left(r + a\right)}$.

\end{itemize}

The grand free energy is taken to be
\begin{equation}
\Phi = \sigma \times \mbox{interface area} + \widetilde{\Delta\mu}
\times \mbox{fluid volume}
\label{simp}
\end{equation}
The stable state can be determined by comparing the free energy of the
different phases. Despite the simplicity of the approach the phase
diagrams in Figures~4 and~5 are reproduced to within 1\%.

Note that a term representing the interaction energy has not been
included explicitly in equation (\ref{simp}). It can be estimated by, for
example, considering only those parts of an interface which adhere to
the substrate as making a contribution. However, for the unbridged
solution the ratio of the surface free energy to the interaction or
volume terms $\sim \sqrt[3]{W_{0}}$. For a bridged phase the volume
component of the free energy increases, but the contribution from the
interactions remains of the same order of magnitude. Therefore, for
the values of $W_{0}$ considered here, the contribution of the
interactions to the total free energy $\sim 1\%$ and can reasonably be
neglected given the level of approximation already inherent in the
approach.

\section{Discussion}

In this paper we have described wetting on a line and rectangular
array of cylinders. For a line of cylinders a capillary condensation
or bridging transition is followed by complete wetting as $\mu
\rightarrow 0$. If the cylinders are sufficiently far apart bridging
does not occur and the wetting transition is suppressed: the cylinders
are now behaving individually rather than as an effectively planar
substrate.

For an array of cylinders we have calculated the phase diagram as a
function of the aspect ratio and reduced chemical potential. Several
different capillary condensation transitions occur: to states bridged
in the $\hat{\vec{\bf x}}$ or $\hat{\vec{\bf y}}$  directions; to a
doubly-bridged phase or to a phase where the liquid completely fills
the volume between the cylinders.

The thin--thick transitions which correspond to wetting on a
cylindrical substrate \cite{GelfandLipowsky,UptonIndekeuYeomans} are
not included in this model where we consider an effective
interface potential with a single minimum. The results of Dobbs and
Yeomans \cite{Harvey2} for adjacent spheres indicate that
including retarded Van der Waals terms in the potential to model such
transitions would not substantially affect capillary condensation,
while the thin--thick transition lines would essentially follow those
for an individual cylinder.

The phase diagrams have been obtained using an interfacial potential
approximation which treats the interface as a sharp delineation between the
two phases. The interface position is obtained as the solution of a
non-linear differential equation. The feasibility of the approach is
dependent on the high symmetry of the system we have considered. To
better model porous media it is important to be able to treat more
complicated substrates. Therefore we have tested a simpler approach
where the shape of the interface in each phase is fed in as an assumption.
The results agree very well with those obtained by the interface
potential method giving us confidence that the simple approach will
give realistic results for substates approximating more closely those
found in porous media.

\samepage{
There have been micromodel experiments undertaken on regular two-dimensional
arrays \cite{Danesh} where capillary condensation is seen to play a
role in the flow of a binary liquid mixture through the model porous
medium. Our results apply to static configurations of the liquid but
may supply insight into the flow properties.}

\section*{Acknowledgements}
It is a pleasure to thank H. T. Dobbs and J. O. Indekeu for many useful
discussions. WRO acknowledges support from a SERC CASE studentship
with British Gas plc, and JMY from a SERC Advanced Fellowship.

\newpage
\section*{Figure Captions}
\begin{description}
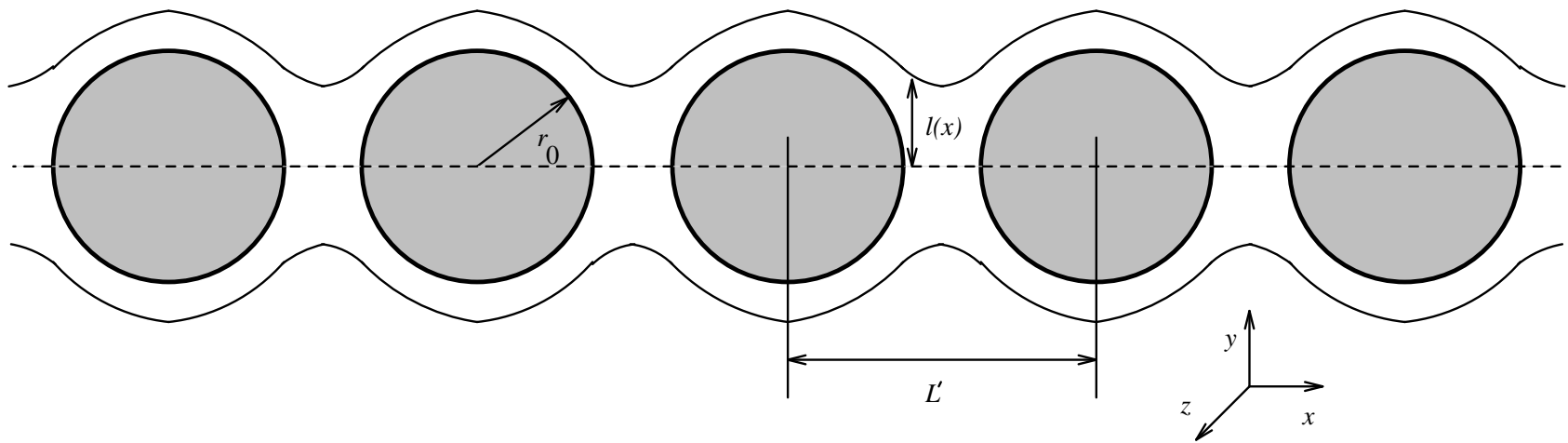
\item[Figure 1]
The arrangement of fluid around a line of cylinders for a
bridged phase.

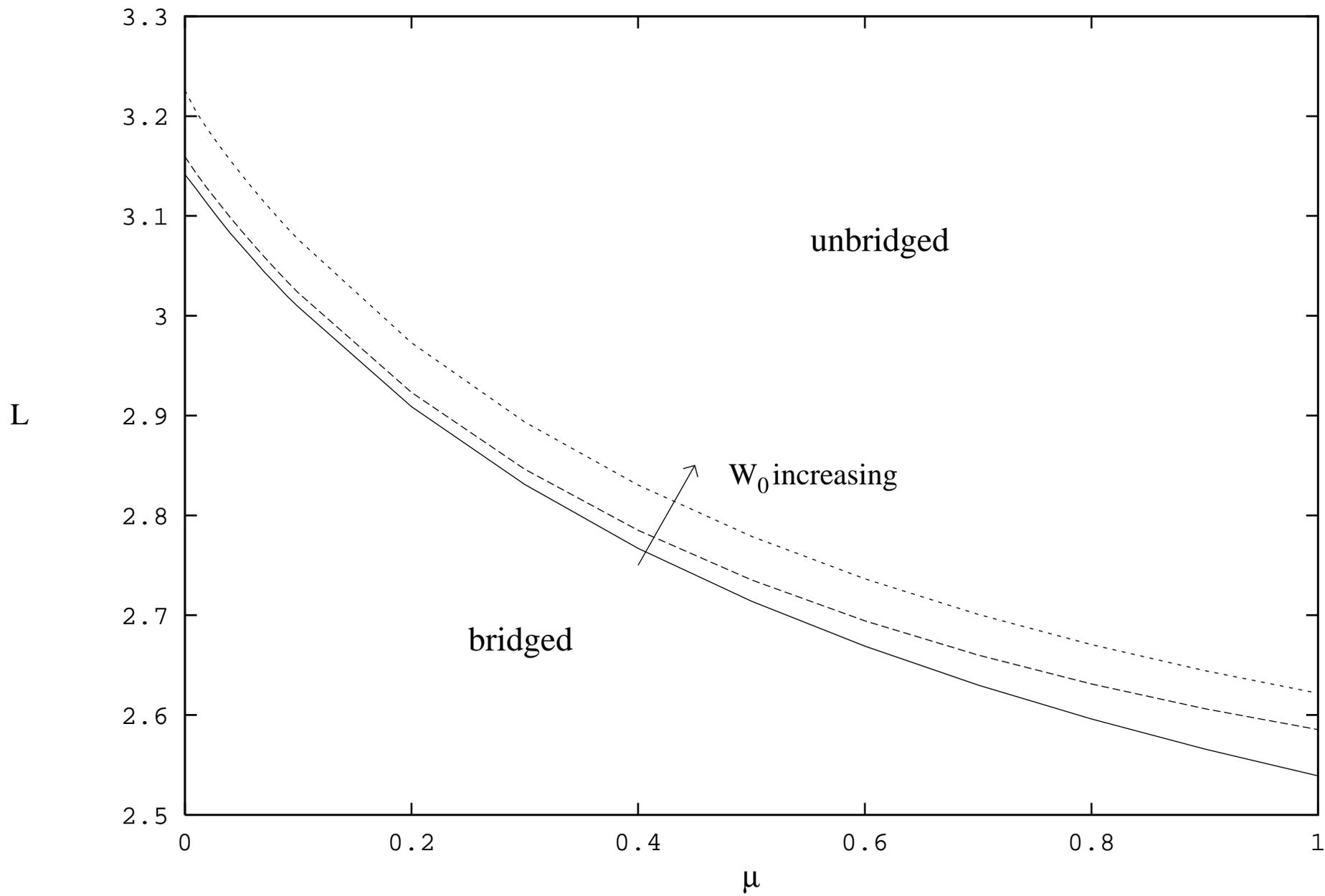
\item[Figure 2]
Dependence of the bridging transition of a single line
of cylinders on the reduced Van der Waals interaction $W_{0}$: solid
line, $W_{0} = 0$; long-dashed line, $W_{0} = 2.5\times10^{-6}$;
shorter-dashed line, $W_{0} = 2.5\times10^{-4}$.

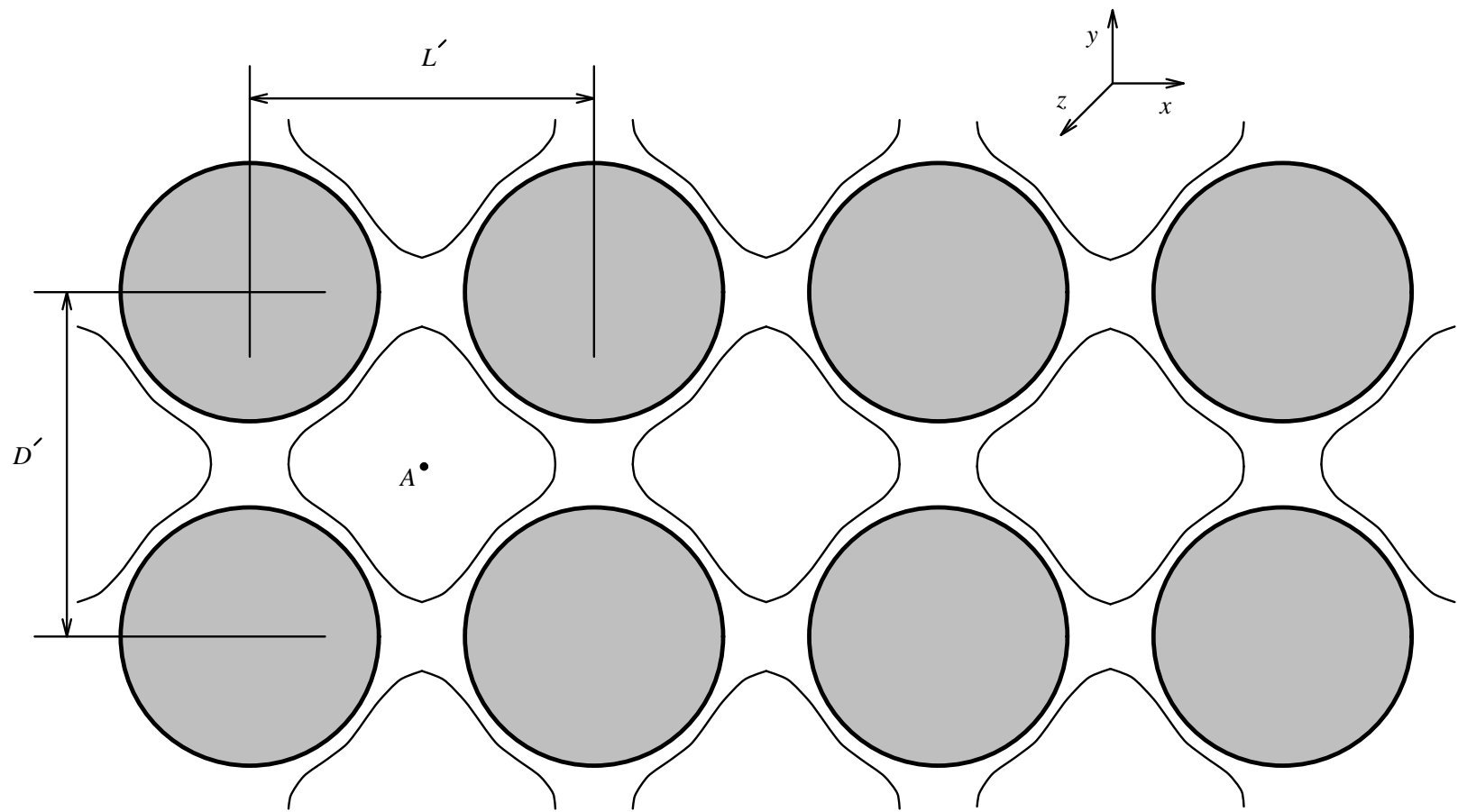
\item[Figure 3]
Part of a rectangular array of cylinders, showing the
doubly-bridged phase.

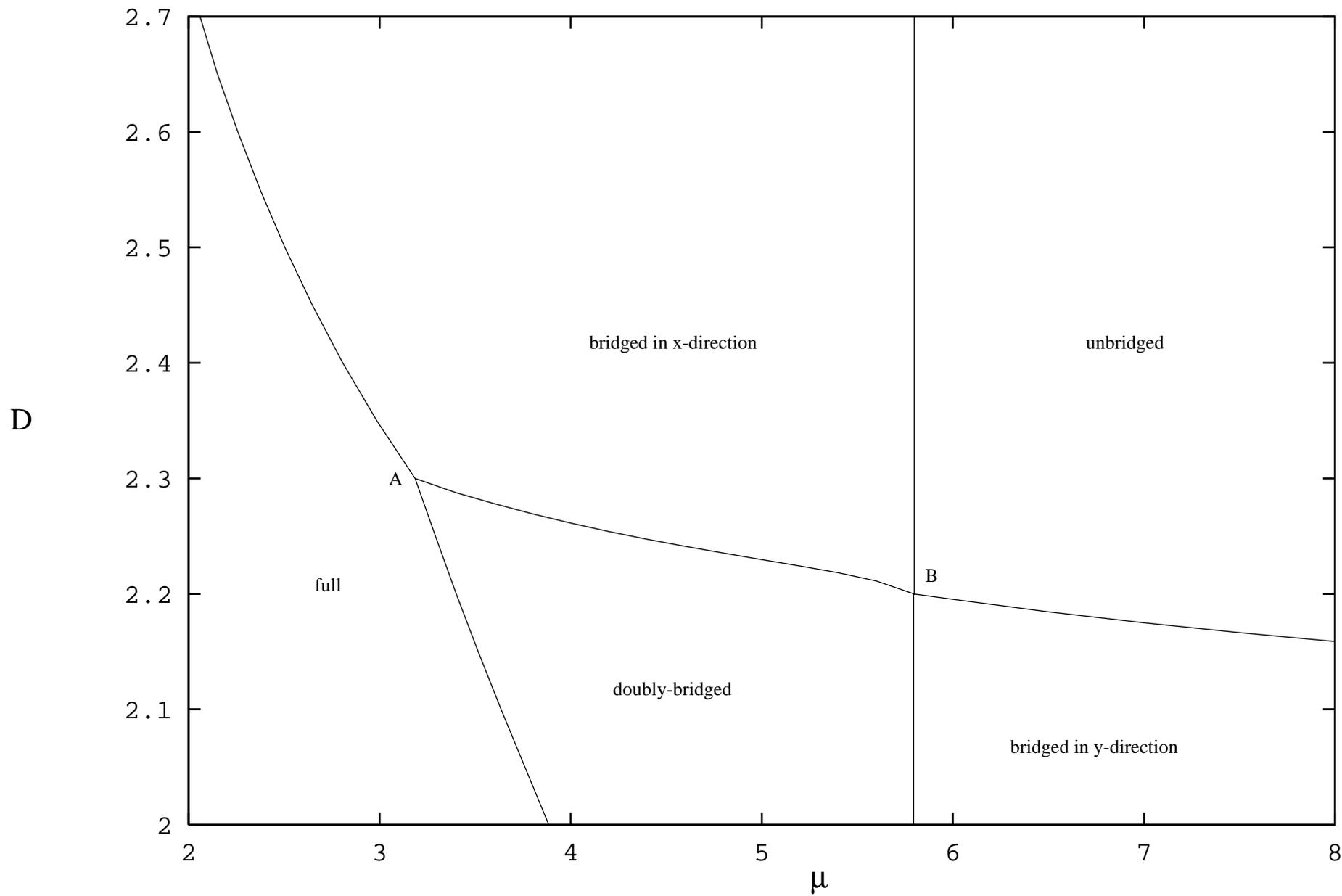
\item[Figure 4]
The phase diagram of an array of cylinders with $L = 2.2$ and
$W_{0} = 2.5\times10^{-6}$.

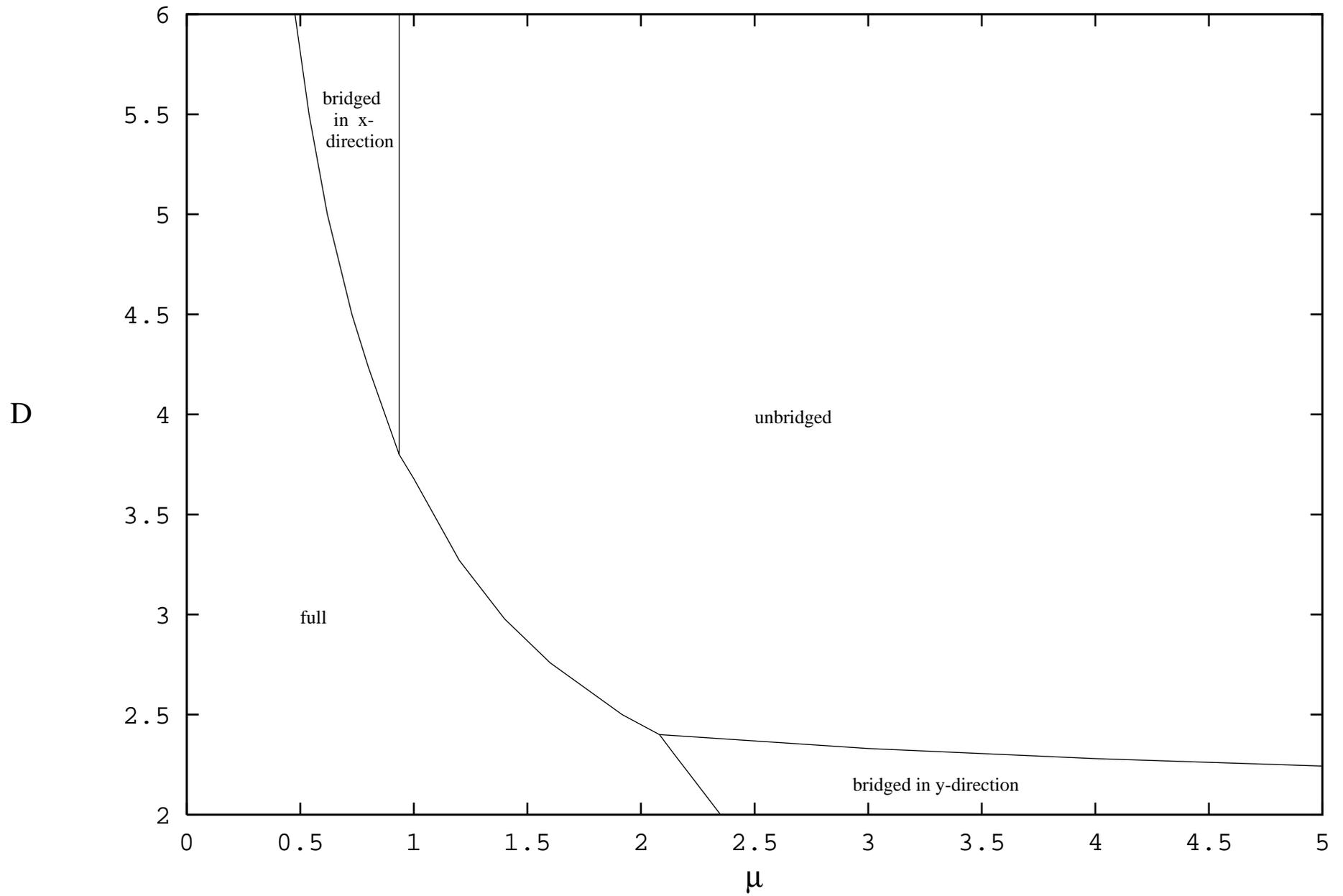
\item[Figure 5]
The phase diagram of an array of cylinders with $L = 2.6$ and
$W_{0} = 2.5\times10^{-6}$.

\item[Figure 6]
The approximate geometries used to model the interface position in the
approach described in \S 4 : (i) unbridged phase, (ii) bridged phase,
(iii) doubly-bridged phase. $a = r_{0}/\mu$.

\end{description}

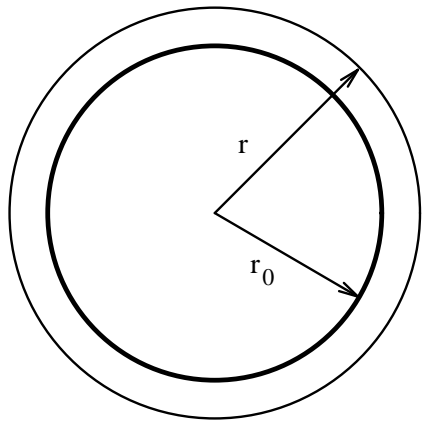
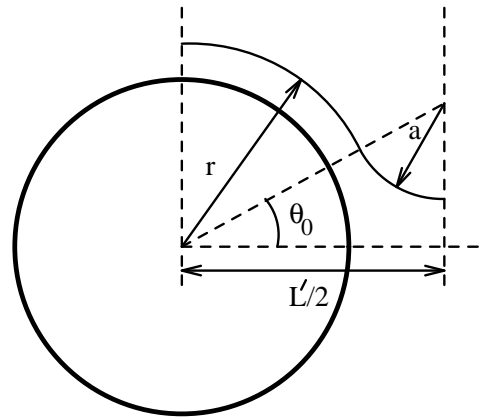
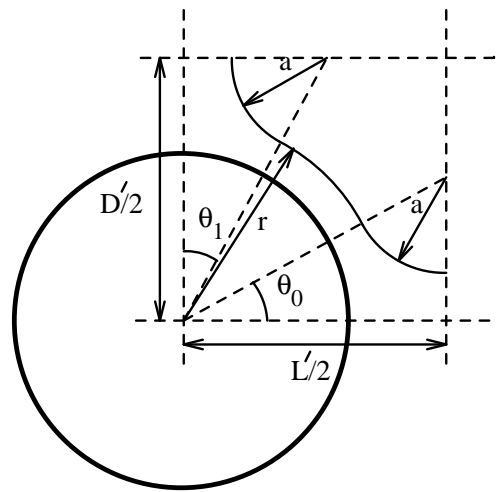
\end{document}